\documentclass[aps,prb,twocolumn,showpacs,floatfix,10pt,superscriptaddress]{revtex4-1}

\usepackage{amsmath}
\usepackage{bm}
\usepackage{graphicx}
\usepackage{amssymb}
\usepackage[caption=false]{subfig}
\usepackage[usenames]{color}
\begin{document}

%
\title{Excitonic Aharonov-Bohm effect in a two-dimensional quantum ring}
\author{C. Gonz\'{a}lez-Santander}
\email{Electronic address: cglezsantander@fis.ucm.es}
\affiliation{Departamento de F\'{i}sica de Materiales, Universidad Complutense de Madrid, E-28040 Spain}
\author{F. Dom\'{i}nguez-Adame}
\affiliation{Departamento de F\'{i}sica de Materiales, Universidad Complutense de Madrid, E-28040 Spain}
\author{R. A. R\"omer}
\affiliation{Department of Physics and Centre for Scientific Computing, University of Warwick, Coventry, CV4 7AL, United Kingdom}

\begin{abstract}
We study theoretically the optical properties of an exciton in a
two-dimensional ring threaded by a magnetic flux. We model the quantum ring by a
confining potential that can be continuously tuned from strictly one-dimensional
to truly two-dimensional with finite radius-to-width ratio. We present an
analytic solution of the problem when the electron-hole interaction is
short-ranged. The oscillatory dependence of the oscillator strength as a function
of the magnetic flux is attributed to the Aharonov-Bohm effect. The amplitude of
the oscillations changes upon increasing the width of the quantum ring. We find
that the Aharonov-Bohm oscillations of the ground state of the exciton decrease
with increasing the width, but remarkably the amplitude remains finite down to
radius-to-width ratios less than unity. We attribute this resilience of the excitonic oscillations to the non-simply connectedness of our chosen confinement potential 
with its centrifugal core at the origin.
\end{abstract}

%

\maketitle


\section{Introduction}

Recent advances in nanofabrication of quantum rings and dots by
self-assembling,~\cite{LorLFK99,LorLGK00,WarSHB00,RibGCM04,TeoCLM10}
lithographic~\cite{BaySHS00,BayKHG03} or etching techniques~\cite{DinALP10} have
opened an active area of research both theoretical and experimental. In such
systems electrons and holes are confined in a small region and consequently the
Coulomb interaction is enhanced. The existence of bound states of electron-hole
pairs offers a unique opportunity to explore the Aharonov-Bohm (AB) effect
~\cite{EhrS49,AhaB59,ByeY61,ChaP93} for excitons in quantum
rings.~\cite{Cha95,RomR00} Despite the exciton being a neutral entity it has
been predicted to be sensitive to a magnetic flux due to its finite size inside
a quantum ring.~\cite{Cha95,RomR00} In experiments, this sensitivity would show
as an oscillatory dependence of both the optical transition energy as well as
the oscillator strength upon the magnetic
flux.~\cite{GalBW02,HafSGWKGSP02,GovKWKU02,BayKHG03,RibGCM04,DinALP10}
Theoretically, the excitonic AB effect has been studied by a variety of
different approaches.  A short-range interaction between the electron and the
hole has been investigated for one-dimensional (1D)
rings,~\cite{Cha95,RomR00,MasMTK01,ShaPR01} where also the effect of an external
electric field can be included.~\cite{FisCPR09} Intermediate models assume
two-dimensional (2D) rings with narrow width under harmonic confinement and
Coulomb-like interaction potentials between the electron and the
hole,~\cite{HuLZX01,HuZLX01} or radial polarized excitons when electrons and
holes move in different circles.~\cite{GovUKW02,DiaUG04,BarPSP06} The excitonic
AB effect in 2D rings has been studied in models with harmonic~\cite{SonU01} and
geometric~\cite{GalBW02,GroGZ06,DaiZ07} confining potential or using a 2D
attractive annular Hubbard model.~\cite{PalDER05,BanCG06} In all cases, the
excitonic AB effect for neutral excitons has been argued to be
suppressed in 2D as the width of the ring is increased.~\cite{LiP11} Recently, experimental
results in molecular-beam epitaxy grown nanorings made by
AsBr$_3$~\cite{DinALP10} etching and on self-assembled InAs/GaAs quantum dots~\cite{TeoCLM10}
report oscillations in the binding energy of neutral excitons which
may be accounted for by the excitonic AB effect.

In this paper, we consider the excitonic AB effect in a confining potential that
can be continuously tuned from strictly 1D to truly 2D with finite
radius-to-width ratio while preserving the central structure of a
ring, namely, its non-simply connectedness due to an infinitely strong repulsion
at the origin.~\cite{TanI96} We present a simple analytic approach to the
excitonic problem when the electron-hole attraction is short-ranged.~\cite{Cha95,RomR00} We then study how the amplitude of the AB oscillations in
the oscillator strength changes upon increasing the width of the ring. We find
that the AB oscillations of the exciton ground state energy decrease with
increasing the width of the quantum ring, but nevertheless the effect remains
noticeable down to regimes with radius-to-width ratios smaller than unity. This
shows the robustness of the  excitonic AB effect in 2D.

\section{Single particle states in the quantum ring}
\label{sec-single}

\begin{figure}[tb]
\includegraphics[width=\columnwidth,clip]{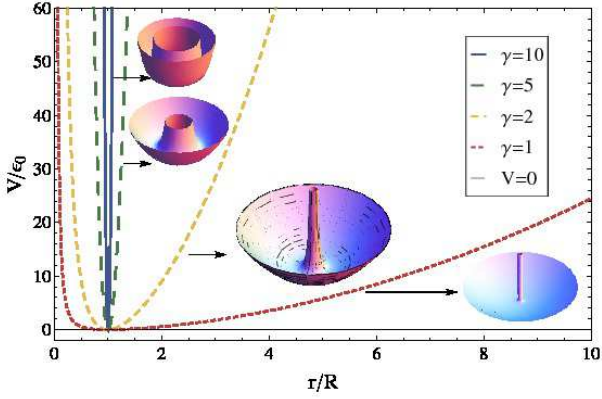}
\caption{Plot of the radial dependence of the confining potential
for $\gamma= 1$ (short-dashed), $2$ (dashed), $5$ (long-dashed) and $10$
(solid). The strongly repulsive core at the origin remains very prominent even for decreasing $\gamma$.}
\label{fig-radialpotential}
\end{figure}

In the absence of Coulomb interaction, the Hamiltonian of a single particle
(electron  or hole) subjected to a magnetic flux in a 2D quantum ring is given
by
\begin{equation}
\mathcal{H}_i=\frac{1}{2m_i}\left(\bm{p}_i-q_i\bm{A}\right)^2+V(r_i)\ ,
\label{eq-singlehamiltonian}
\end{equation}
where $m_i$, ${\bm p}_i$ and ${\bm A}$ are the effective mass, the momentum in
the plane and magnetic vector potential, respectively. Here the subscript
$i=e,h$ refers to the electron and the hole, respectively. Electric charges are
$q_e=-e$ and $q_h=e$. The quantum ring is modeled by an anharmonic, axially
symmetric potential with a centrifugal core~\cite{TanI96,KovC09}
\begin{equation}
V(r_i)=\frac{V_0}{2}\left[\frac{R^2}{r_i^2}+\frac{r_i^2}{R^2}\right]-V_0\ .
\label{eq-confiningpotential}
\end{equation}
The confining potential has the minimum at $|{\bm r}_i|=R$  (see
Fig.~\ref{eq-singlehamiltonian}) and, for this reason, $R$  will be used as a
convenient measure of the effective ring radius. Close to the minimum the
potential reduces to the well-known displaced parabola, $V(r_i)=2V_0(r_i/R-1)^2
\equiv (1/2)m_i\omega^2(r_i-R)^2$, used in other theoretical studies of 2D
quantum rings.~\cite{SonU01} As $r_{i}\rightarrow 0$, we see from \eqref{eq-confiningpotential} that the centrifugal core 
assures the survival of the essential feature of a ring: its repulsive barrier in the center. The effective width $W$ of the quantum ring
can be estimated from the  single-particle ground state in the harmonic
potential, namely  $W=(\epsilon_0/2 V_0)^{1/4} R$, where
$\epsilon_0=\hbar^2/2mR^2$ is the ring-size quantization energy.\cite{TanI96} Notice that we
assume that $W$ is the same for electrons and holes, namely $m_e=m_h\equiv m$.
For the purpose of this work, all energies will be measured in units of
$\epsilon_0$ and we parametrize the strength of the confining potential by the
radius-to-width ratio  $\gamma \equiv R/W= (2 V_0/\epsilon_0)^{1/4}$. When
$\gamma \rightarrow \infty$, we approach the limit of a 1D ring,
whereas $\gamma\rightarrow 0$ corresponds to an anti-dot geometry.~\cite{TanI96}
Figure~\ref{fig-radialpotential} shows the radial confining potential for
different values of $\gamma$.

In order to study the AB effect in the quantum ring, we choose
$\bm{A}\equiv (A_r,A_\theta)=(0,\Phi\, h/e 2\pi r)$, corresponding to an infinitely thin magnetic flux
piercing the plane of the ring perpendicularly.
Here $\Phi$ is the dimensionless flux through the ring and $h/e$ the universal flux quantum.%
\cite{Note1}
We note that due to the axial symmetry aroud the ring axis, 
all our results for energies have to be periodic in $\Phi$ with period $1$ and we hence restrict ourselves to the sector $\Phi\in [0,1]$.
Then the Schr\"{o}dinger equation for
the electron in polar coordinates $\bm{r}_e=$($r_e$,$\theta_e$) is
written in dimensionless form
\begin{eqnarray}
\lefteqn{\mathcal{H}_e \psi_{M_e}(\bm{r}_e) = \lambda_{M_e}
\psi_{M_e} (\bm{r}_e) =} \nonumber \\
& = &  \big[-\frac{\partial^2}{\partial \rho_e^2}-\frac{1}{\rho_e}
\frac{\partial}{\partial \rho_e}-
\frac{1}{\rho_e^2}\frac{\partial^2}{\partial \theta_e^2} \nonumber \\
 & & - \frac{2i\Phi}{\rho_e^2} \frac{\partial}{\partial\theta_e}
 +\frac{\Phi^2}{\rho_e^2} + \frac{V_e}{\epsilon_0}\big]
\psi_{M_e} (\bm{r}_e) \ ,
\label{eq-dhamiltonian}
\end{eqnarray}
where $M_e=(n_e,\ell_e)$ represents the set of quantum numbers for the electron,
which are $n_e=0,1,2,\ldots$ and $\ell_e=0,\pm 1\,\pm 2,\ldots$ For brevity we
define the dimensionless energy $\lambda_{M_e}=E_{M_e}/\epsilon_0$ and
radial coordinate $\rho_e=r_e/R$.  The Schr\"{o}dinger equation for the hole is the same aside
from a change in sign in the linear term on $\Phi$, and with a set of quantum
numbers $M_h=(n_h,\ell_h)$.

The normalized eigenfunctions of Eq.~\eqref{eq-dhamiltonian} are given
by~\cite{TanI96}
\begin{subequations}
\begin{eqnarray}
\psi_{M_e}(\bm{r}_e) &=&\frac{e^{-i\ell_e\theta_e}}{\sqrt{2\pi}}\,
\mathcal{R}_{M_e}(r_e) \ , \\
\mathcal{R}_{M_e}(r_e) &=& \frac{1}{R}
\left[\frac{\Gamma(n_e+1)}{2^{k_e}\Gamma(n_e+k_e+1)}\right]^{1/2}\nonumber \\ &
\times & (\rho_e \gamma)^{k_e} e^{-\rho_e^2\gamma^2/4} L_{n_e}^{k_e}
\left(\frac{\rho_e^2\gamma^2}{2}\right) \ ,
\label{eq-singlewave}
\end{eqnarray}
\end{subequations}
where $k_e=\sqrt{f_e^2+\gamma^4/4}$ with $f_e=\ell_e-\Phi$ defining an effective
angular quantum number due to the confinement and the magnetic flux. $L_n^k$
stands for the generalized Laguerre polynomials. The corresponding dimensionless
energies are $\lambda_{M_e}=\gamma^2(2n_e+1+k_e)-\gamma^4/2 $. The
eigenfunctions and energies for the hole are the same as for the electron, with
a effective angular quantum number $f_h=\ell_h+\Phi$. The dimensionless zero point energy ($n_e=\ell_e=\Phi=0$) for the electron is $\lambda_e^0=\gamma^2$.

\begin{figure}[tbh]
\includegraphics[width=\columnwidth,clip]{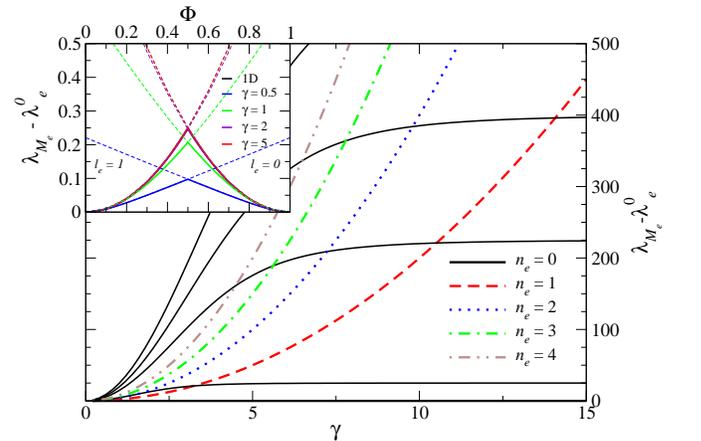}
\caption{Dimensionless electron energy $\lambda_{M_e}$ as function of the
radius-to-width ratio $\gamma$. Solid, dashed, dotted, dotted-dashed and
double-dotted-dashed lines corresponds to $n_e=0, 1, 2, 3, 4$ and $l_e=0$. For $n_e=0$ we also show $l_e=5, 10, 15, 20$. The flux is $\Phi=0$ in all cases. The inset
shows the flux dependence of the energy of the ground state for different
$\gamma$. The black lines corresponds to 1D results.}
\label{fig-singlenergies}
\end{figure}

Figure~\ref{fig-singlenergies} shows the electron energy as a function of the
parameter $\gamma$.  We note that levels at higher values of the  quantum number
$n_e$ become increasingly uncoupled for $\gamma > 5$ and hence we expect to see
nearly 1D behavior for $\gamma$ values beyond this regime.~\cite{RomR00} From
the inset is clearly observed that in this 2D confinement regime, in the absence
of interaction, the ground state energy for the electron (once sustracted the
zero point energy) describes an oscillation with the magnetic flux.  For
$\gamma > 5$, the 2D oscillation is indistinguishable from the 1D case.

\section{Solution of the excitonic case}
\label{sec-exciton}

Within the effective-mass approximation, the Hamiltonian of the  interacting
electron-hole pair is given by
$\mathcal{H}=\mathcal{H}_e+\mathcal{H}_h+\mathcal{H}_{e-h}$, where
$\mathcal{H}_{e-h}$ is the interaction term.
We model the excitonic interaction between the electron and the hole as a
short-range potential of the form
$
\mathcal{H}_{e-h}(\bm{r}_e,\bm{r}_h)/\epsilon_0=(2\pi)^{3/2} v_0 RW\delta(\bm{r}_e-\bm{r}_h)
$, 
where $v_0<0$ parametrizes the attractive interaction strength.
This contact interaction is the same used in Refs.~\onlinecite{RomR00,FisCPR09} extended
to a 2D case, where the area of the ring is $2\pi RW$.
In this definition we have carefully chosen the prefactors such that in the 1D limit, $\gamma\to\infty$,
the values of $v_0$ become identical to the corresponding 1D parameter and facilitate comparison  with the results of~Refs.\ \onlinecite{RomR00,FisCPR09}.
Thus, we express $v_0$ as $-\alpha/\pi^2$ where $\alpha$ denotes the ratio of 1D excitonic Bohr radius to ring circumference.~\cite{RomR00}

Before continuing with the detailed study of the model, let us discuss some of the assumptions made and the limitations which we will encounter. 
Let us first emphasise that the restriction to equal electron and hole masses is simply a presentational convenience; all calculations shown here can easily be generalised to the case of unequal masses\cite{FisCPR09} but with a certain loss of clarity in the mathematical expressions. Nevertheless, we shall present some results for unequal masses later. 
The assumption of an infinitely thin current-carrying solenoid generating the
magnetic flux $\Phi$ is a theoretical construct. The experiments cited in the
introduction all use a magnetic field $B$ to generate the required $\Phi$. This
results in an additional, diamagnetic term proportional to $B^2$, which we
ignore here similarly to the experimental
papers.\cite{RibGCM04,BayKHG03,DinALP10,GalBW02,HafSGWKGSP02,GovKWKU02} 
Certainly the most drastic assumption seems to be the $\delta$-function
potential for the two-particle interaction. Its use is of course motivated by
our resulting ability to reduce the computational difficulties as we will show
below. Nevertheless, we wish to emphasise that there are also certain conceptual
advantages associated with it: (i) in 1D, the $\delta$-function interacting
many-particle problem has been solved exactly and hence the expression for the
exciton binding energy on a line is known in terms of $v_0$.\cite{LieL63} (ii)
In Ref.~\onlinecite{RomR00}, it was shown how the Bohr radius of the exciton
similarly depends on $v_0$. Both these parameters will of course vary when
another form of interaction is considered. However, as also shown in Ref.\
\onlinecite{RomR00}, it is the ratio $\alpha$ introduced above which governs the
strength of the AB oscillations. The effect of other two-particle interaction
potentials along the ring, when expressed in terms of $\alpha$, will lead to
similar AB oscillations and we expect at least qualitative agreement. Even for a
long-range potential such as the Coulomb interaction, we expect this to hold as
long as the overlap of wave packets on opposite sides of the ring, i.e.\ across
the origin at $\bm{r}=0$, can be neglected. For the confining potential
considered here, with its strong centrifugal core, this should be a rather good
approximation.

We construct the exciton eigenfunction as a linear combination of the electron
and hole single-particle eigenfunctions
\begin{equation}
\Psi(\bm{r}_e,\bm{r}_h)= \sum_{M_eM_h}A_{M_eM_h}\psi_{M_e}(\bm{r}_e)\psi_{M_h}(\bm{r}_h) \ .
\label{eq-excitonwave}
\end{equation}
The Schr\"odinger equation for the electron-hole pair may now be cast in
equivalent form
\begin{equation}
 \begin{split}
  \sum_{M_eM_h}A_{M_eM_h}(\lambda_{M_e}+\lambda_{M_h}-\Delta)\psi_{M_e}(\bm{r}_e)\psi_{M_h}(\bm{r}_h)  \\
+ (2\pi)^{3/2} v_0 RW\delta(\bm{r}_e-\bm{r}_h) \Psi(\bm{r}_e,\bm{r}_h)=0 \ ,
 \end{split}
\label{eq-A1}
\end{equation}
where $\Delta$ is the excitonic energy in units of $\epsilon_0$. Following an analogous procedure as in Ref.\ \onlinecite{FisCPR09} the coefficients
$A_{M_eM_h}$ are obtained multiplying Eq.~\eqref{eq-A1} by $\psi_{M_e}^{\dagger}(\bm{r}_e)\psi_{M_h}^{\dagger}(\bm{r}_h)$
and integrating over the coordinates
\begin{equation}
A_{M_eM_h} = -\frac{(2\pi)^{3/2} v_0 RW}{\lambda_{M_e}+\lambda_{M_h}-\Delta}\,
G_{M_eM_h} \ ,
\label{eq-A2}
\end{equation}
where we have defined
\begin{equation}
G_{M_eM_h}=\int d^2 \bm{r} \Psi(\bm{r},\bm{r})\,\psi^{\dagger}_{M_e}(\bm{r})
\psi^{\dagger}_{M_h}(\bm{r}) \ .
\label{eq-A3}
\end{equation}
Setting $\bm{r}_e = \bm{r}_h = \bm{r}$ in the expansion of
~\eqref{eq-excitonwave}, multiplying by
$\psi_{M'_e}^{\dagger}(\bm{r})\psi_{M'_h}^{\dagger}(\bm{r})$  and integrating
over the coordinates we finally obtain
\begin{equation}
G_{M'_eM'_h}=\sum_{M_eM_h}G_{M_eM_h}P_{M_eM_hM'_eM'_h}(\Delta) \ ,
\label{eq-matrix}
\end{equation}
with
\begin{equation}
 \begin{split}
 & P_{M_eM_hM'_eM'_h} =-\frac{(2\pi)^{3/2} v_0 RW}{\lambda_{M_e}
   +\lambda_{M_h}-\Delta}  \\
 &\times  \int d^2 \bm{r} \, \psi_{M_e}(\bm{r})\psi_{M_h}(\bm{r})
  \psi^{\dagger}_{M'_e}(\bm{r})\,\psi^{\dagger}_{M'_h}(\bm{r})\ .
  \label{eq-A5}
  \end{split}
\end{equation}

To proceed we define the total angular momentum of the electron-hole pair in
units of $\hbar$ as $L=\ell_e+\ell_h$. Because the system is axially symmetric,
only states with $L=L'$ can contribute to the excitonic system. This condition
is even more restrictive under the dipole approximation, i.e.\ only excitons with
total angular momentum $L=0$ can absorb light polarized perpendicular to the ring. Therefore Eq.~\eqref{eq-A5}
reduces to
\begin{equation}
 \begin{split}
&  P_{M_eM_hM'_eM'_h} =-\frac{\sqrt{2\pi}v_0 RW}{\lambda_{M_e}+\lambda_{M_h}
  -\Delta}  \\
&\times  \int_{0}^{\infty} d r \,r \, \mathcal{R}_{M_e}(r)
 \mathcal{R}_{M_h}(r)\mathcal{R}_{M'_e}(r)\mathcal{R}_{M'_h}(r)\ .
\label{eq-Pmmmm}
 \end{split}
\end{equation}
We note that in the limit $\gamma\to\infty$, the integrals in \eqref{eq-Pmmmm}
reduce to $1/\sqrt{2 \pi}RW$ for $n_e=n_h=n'_e=n'_h=0$ and
$l_e+l_h=l'_e+l'_h$. For other combinations of $n_e$, $n_h$, $n'_e$, and $n'_h$, the corresponding $P_{M_eM_hM'_eM'_h}$ are less important due to the energy denominator in \eqref{eq-Pmmmm}.

Similarly as for 1D quantum
rings,~\cite{RomR00,FisCPR09} there is no analytical solution of
Eq.~\eqref{eq-matrix} for finite values of $v_0$. In order to find approximate
solutions, we hence need to cut off the sums at some maximally allowed
values for $M_e$ and $M_h$. 
Figure~\ref{fig-singlenergies} shows that for smaller values of $\gamma$, i.e.\
increasing ring width $W$, the level separation between the quantum
states of the single particles is decreased. Therefore we use different
$\ell_\mathrm{max}$ and $n_\mathrm{max}$ values depending on our choice of
$\gamma$. We have tested that our results do not change appreciably for the
range of $\Phi$ and $v_0$ considered here. As in Ref.~\onlinecite{FisCPR09},
Eq.~\eqref{eq-matrix} is reformulated as a standard left-eigenvalue equation
$G_{K'}=\sum_{K}G_{K}P_{KK'}(\Delta)$ after mapping the quantum numbers $M_e,M_h\to
K$ and $M'_e,M'_h\to K'$ according to
$K=(\ell+\ell_\mathrm{max})(n_\mathrm{max}+1)^2+n_e(n_\mathrm{max}+1)+n_h+1$ such that $K, K'=1, 2, \ldots, (1+2 \ell_\mathrm{max})(1+n_\mathrm{max})^2$. The excitonic
energies are obtained numerically by determining the values of $\Delta$ which
result in the matrix $P_{KK'}$ having an eigenvalue equal to 1. For a given $\Delta$, all
eigenstates can be found using \eqref{eq-matrix}, \eqref{eq-A2} and
\eqref{eq-excitonwave}. An advantage of our approach is that it allows us to target the ground state directly by choosing a suitable starting value for $\Delta$.

\section{Results}

\begin{figure}[tb]
\includegraphics[width=\columnwidth,clip]{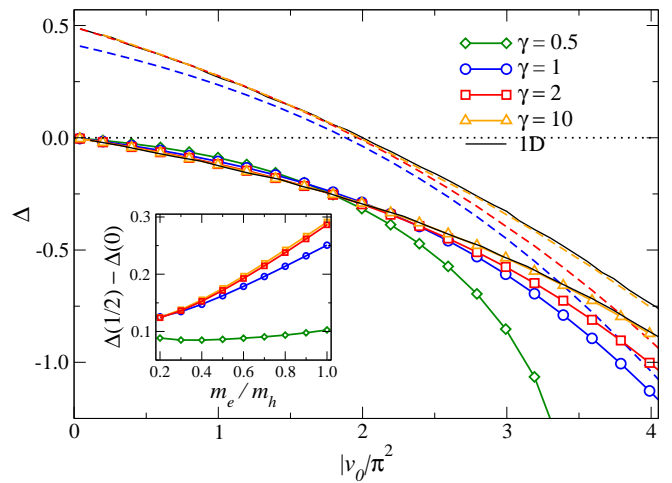}
\caption{Exciton energy $\Delta$ at magnetic flux $\Phi=0$ (solid lines) and
$1/2$ (dashed lines) plotted as a function of the interaction strength $v_0$.
For clarity,  symbols are shown for $\Phi= 0$ only and the results for
$\gamma=0.5$ at $\Phi=1/2$ have been suppressed. The thin dotted horizontal line
denotes the onset of the single-particle  continuum at $\Phi=0$. The two thin
black lines denote the 1D limit for  $\Phi=0, 1/2$. The inset shows
the amplitude of the excitonic AB oscillations as a function of the ratio
$m_e/m_h$ at interaction strength $v_0=-2/\pi^2$.}
\label{fig-energystrength}
\end{figure}

In Fig.~\ref{fig-energystrength} we plot the ground state energy $\Delta$ defined by
Eq.~\eqref{eq-matrix} with $\ell_\mathrm{max}=40$ for $\gamma>0.5$ and $n_\mathrm{max}=5$ ($K_\mathrm{max}=2916$) for $\gamma<5$ or $n_\mathrm{max}=2$ ($K_\mathrm{max}=729$) for $\gamma>5$ and
as a function of $v_0$ for different values of $\gamma$. For $\gamma=0.5$ we have used $\ell_\mathrm{max}=30$ and $n_\mathrm{max}=6$ ($K_\mathrm{max}=2989$). 
Here and in all following figures, when plotting the excitonic energies $\Delta$,
we have subtracted the zero-point energy  $2\gamma^2$ of the non-interacting electron-hole system.

We see that for all $\gamma$ and $\Phi$ values, the increase of the interaction
strength $v_0$ leads to the formation of a state with decreasing energy values
below the onset of the free-particle continuum. We also compare in
Fig.~\ref{fig-energystrength} the 2D exciton results with the 1D ring
studied in Refs.~\onlinecite{RomR00,FisCPR09}.  When the radius of the ring is $10$
times its width ($\gamma=10$) the 2D excitonic behavior is essentially
indistinguishable  from the 1D results in the range of $v_0$ values studied. In
particular, the differences between energies at different flux values at large
$\gamma$ decrease. Nevertheless, for small $\gamma \lesssim 3$ different
magnetic flux values lead to quite distinct $\Delta$ values  --- even in a 2D
quantum ring the exciton is sensitive to the magnetic flux.  It is also
interesting to note that for large $\gamma$, the bound state energies are more
negative for larger values of $v_0$~[\onlinecite{RomR00}] whereas for $\gamma\lesssim 1$
we find evidence that smaller $\gamma$ values lead to smaller differences
between different values of $\Phi$.

The above results have been obtained assuming the same effective mass
for the electron and the hole. To addrees the question of the  robustness of the
excitonic AB effect in a more realistic situation with different effectives
electron and hole masses, we have calculated the exciton energy as a function of
the ratio $m_e/m_h$. Let us define the amplitude of the excitonic AB
oscillations as the difference of the exciton energy at $\Phi=1/2$ and $\Phi=0$,
namely $\Delta(1/2)-\Delta(0)$. The inset of Fig.~\ref{fig-energystrength} shows
this amplitude  as a function of the ratio $m_e/m_h$ for different values of the
radius-to-width ratio $\gamma$ at interaction strength $v_0=-2/\pi^2$. In 2D
rings ($\gamma=0.5$) the energy difference is almost constant and the assumption
of equal masses is well justified. Upon approaching the 1D limit, i.e.
increasing $\gamma$, the  amplitude of excitonic AB oscillations is reduced but
the effect is still revealed. As an example, in common III-V compound
semiconductors the ratio of the electron and ligh hole masses typically ranges
from $0.6$ to $0.9$, and it can be seen in the inset of
Fig.~\ref{fig-energystrength} that the reduction of the amplitude is small.

\begin{figure}[tb]
\centerline{\includegraphics[width=\columnwidth,clip]{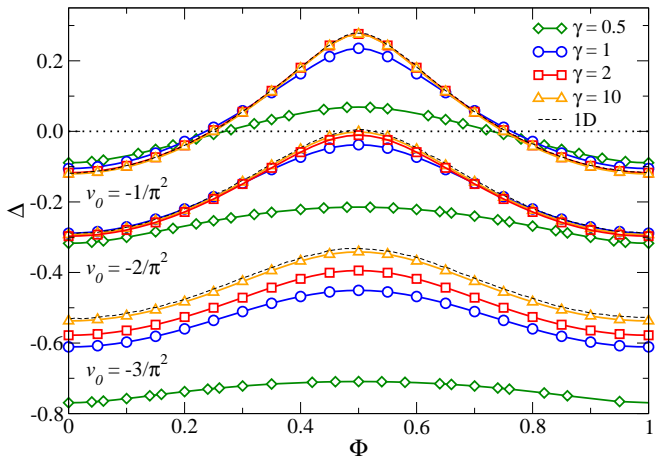}}
\caption{Exciton energy $\Delta$ as function of the magnetic flux $\Phi$ for different values of interaction strength $v_0$ and radius-to-width ratio $\gamma$.
The thin dotted horizontal line denotes the onset of the single-particle continuum at $\Phi=0$. Only every second data point is shown for clarity in each curve.}
\label{fig-energyflux_all}
\end{figure}

Figure~\ref{fig-energyflux_all} shows the AB oscillations of the exciton energy
as a function of the magnetic flux $\Phi$ within one flux period at different
values of $\gamma$ and $v_0$.
In agreement with Fig.~\ref{fig-energystrength}, we find that the AB
oscillations are retained  for radius-to-width ratios ranging from $\gamma=0.5$
to $10$.  This shows that the excitonic AB effect remains robust even in
a ring of finite width. Upon increasing the $\gamma$ values for different $\Phi$
values, we find mostly a moderate increase of the exciton energy, except in the
vicinity of $\Phi= 0.5$ where even the reverse tendency can be observed.

\begin{figure}[htb]
\centerline{\includegraphics[width=\columnwidth,clip]{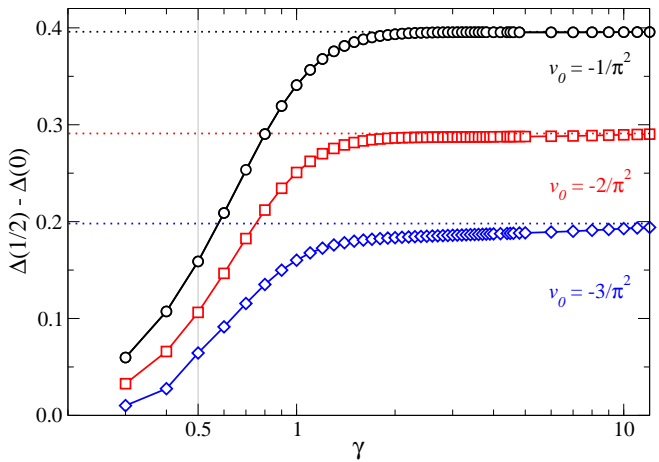}}
\caption{Amplitude of the AB oscillations $\Delta(1/2)-\Delta(0)$ as a function of radius-to-width ratio
 $\gamma$ for different interaction strength $v_0$. The dashed horizontal lines correspond to the 1D limit~\cite{RomR00}, the vertical line denotes the $\gamma=0.5$ values.}
\label{fig-energyflux_2}
\end{figure}

In Fig.~\ref{fig-energyflux_2} we plot the amplitude of the excitonic AB oscillations for different interaction strength $v_0=-1/\pi^2,-2/\pi^2,-3/\pi^2$ as $\gamma$ is varied.
We see that upon decreasing $\gamma$ from the nearly 1D behaviour at $\gamma=10$
towards $\gamma\approx 1.5$, there is only a slight decrease in the amplitude of the
AB oscillations.  Upon further decreasing $\gamma$, the oscillations weaken more rapidly, but
even at $\gamma=0.5$, they retain about $30$--$40\%$ of their original value.
Results for other values of $v_0$ are similar. This again shows that even for
rather wide rings, the excitonic AB oscillations persist in this 2D case.

\begin{figure*}[htb]
  \centering
   \subfloat[$\gamma=1$]{\label{fig-WF-g1}\includegraphics[width=0.5\columnwidth,clip]{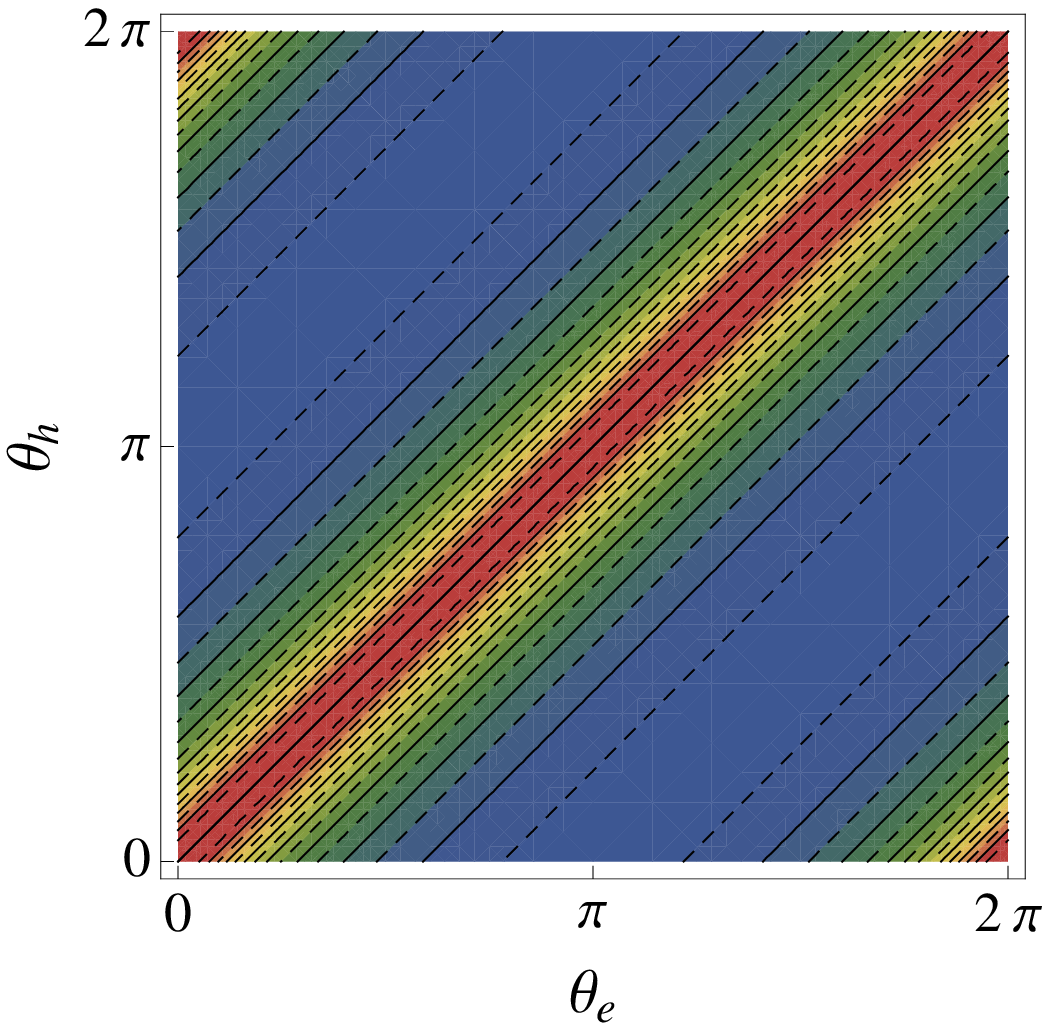}}
  \subfloat[$\gamma=1$]{\label{fig-WFR-g1}\includegraphics[width=0.5\columnwidth,clip]{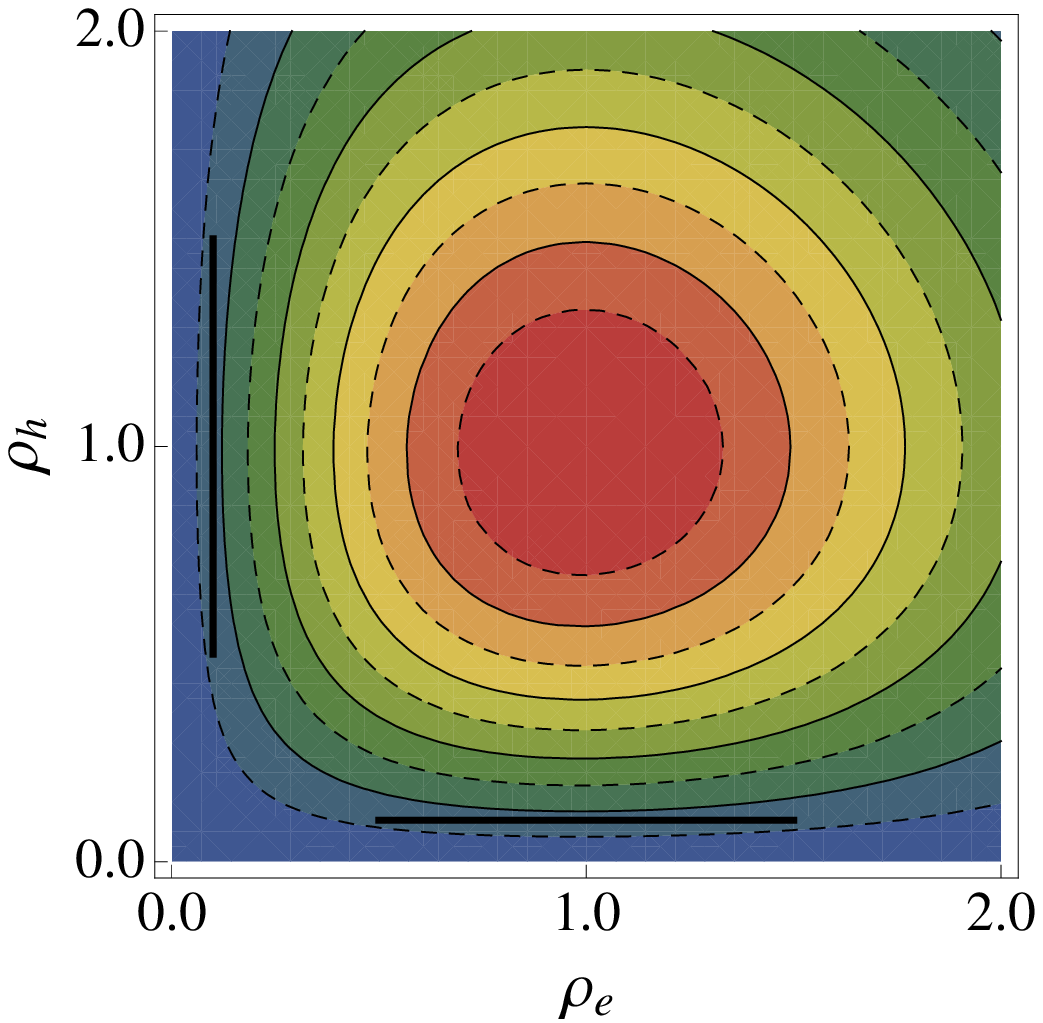}}
\subfloat[$\gamma=10$]{\label{fig-WF-g10}\includegraphics[width=0.5\columnwidth,clip]{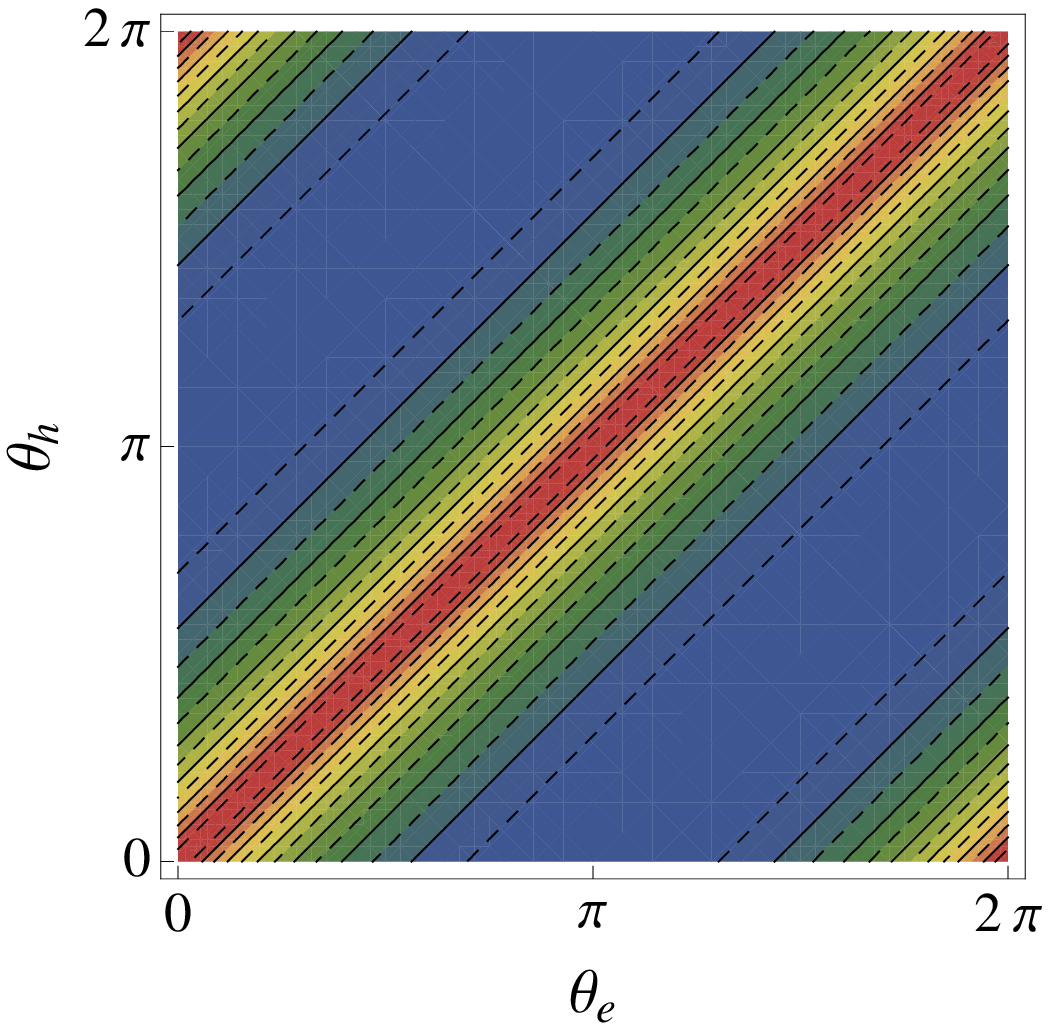}}
  \subfloat[$\gamma=10$]{\label{fig-WFR-g10}\includegraphics[width=0.5\columnwidth,clip]{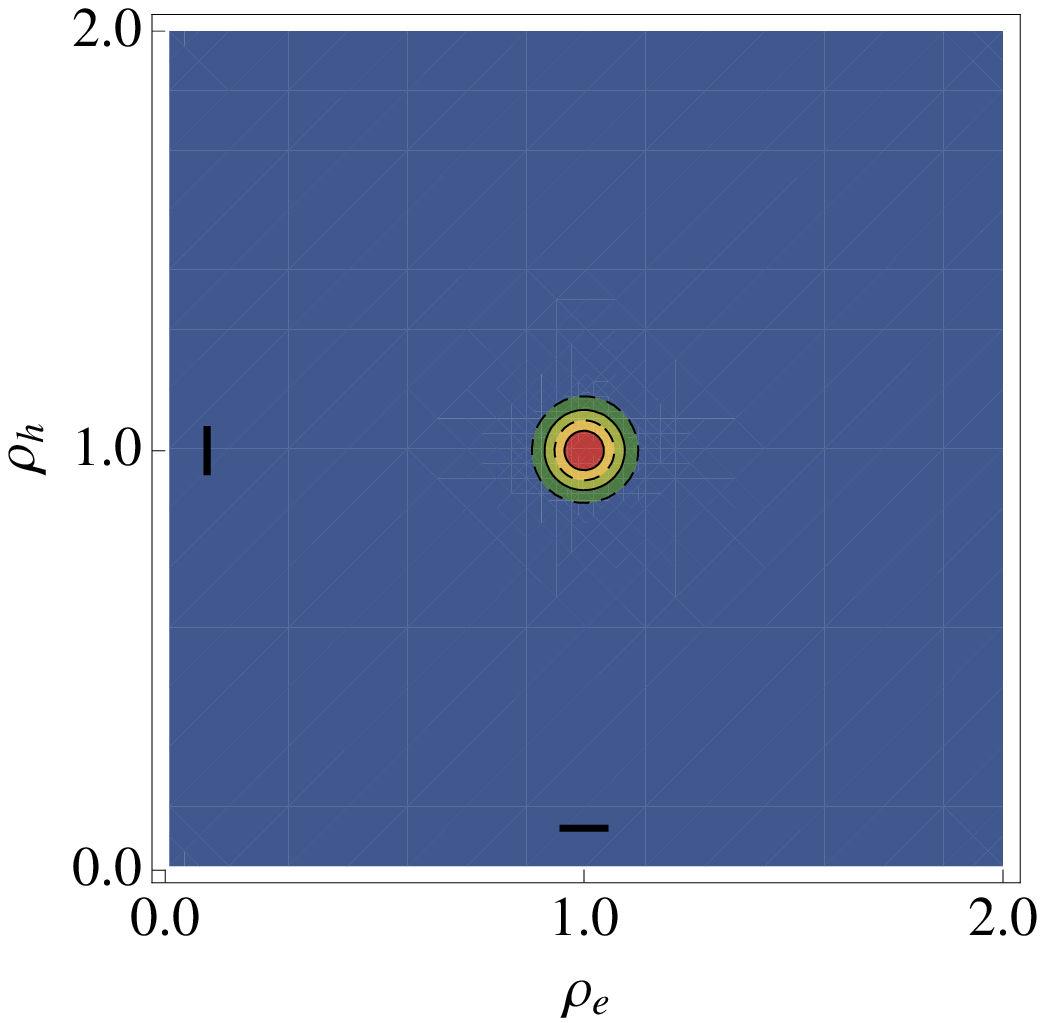}}
  \caption{Dependence of the {\em local} wavefunction probability $|\Psi|^2$ on
(a,c) angular coordinates $\theta_e$, $\theta_h$ and (b,d) radial coordinates $\rho_e=r_e/R$, $\rho_h=r_h/R$ for $\gamma=1$ (a,b), and $10$ (c,d)
for $v_0=-2/\pi^2$ and $\Phi=0$.
The values of $|\Psi(\theta_e,\theta_h)|^2$ and $|\Psi(\rho_e,\rho_h)|^2$ have been normalized to lie in $[0,1]$.
The colors go from $|\Psi|^2\in\ [0.9,1]$ (red) to $|\Psi|^2\in\ [0, 0.1]$ (blue) in steps of $0.05$ for (a,c), $0.2$ for (b) and $0.1$ for (d).
The thick black lines indicate in (b) and (d) the width $W$ of the ring in each case.}
  \label{fig-WF}
\end{figure*}

In Fig.~\ref{fig-WF} we plot the exciton probablity density
$|\Psi(\bm{r}_e,\bm{r}_h)|^2$ for different values of $\gamma$.  We integrate
$|\Psi|^2$ over the radial coordinates $\rho_e$, $\rho_h$ and hence retain the angular dependence
in Figs.~\ref{fig-WF-g1} and~\ref{fig-WF-g10}, whereas in
Figs.~\ref{fig-WFR-g1} and~\ref{fig-WFR-g10} we integrate out the angular
degrees of freedom and retain the $\rho_e$, $\rho_h$ dependence.  From these figures we
conclude that the exciton fills the available width of the ring.  Figures~\ref{fig-WF-g1} and~\ref{fig-WF-g10} show that the
exciton is indeed bound, i.e.\ the majority of the weight of $|\Psi|^2$
resides along the diagonal $\theta_e=\theta_h$. Analogous results are obtained for different $\Phi$ and $v_0$. This is similar to the 1D
behavior described in Refs.~\onlinecite{RomR00,FisCPR09}.

\begin{figure}[ht]
\centerline{\includegraphics[width=\columnwidth,clip]{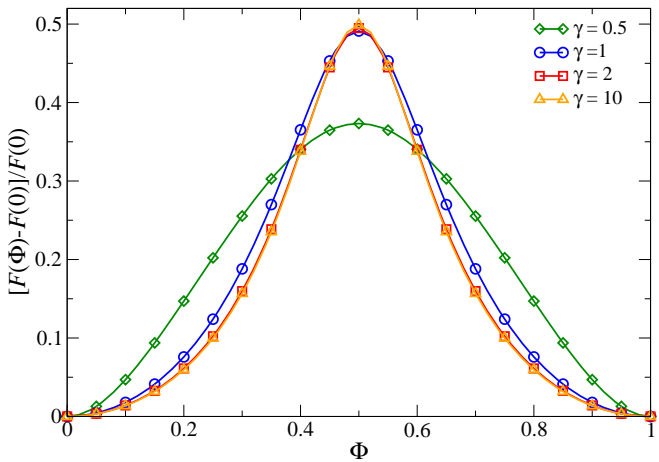}}
\caption{Normalized oscillator strength $[F(\Phi)-F(0)]/F(0)$ as a function of magnetic flux $\Phi$ for $\gamma= 0.5, 1, 2, 5$ and $10$ at interaction strength $v_0=-2/\pi^2$. Only every second data point is shown for clarity.}
\label{fig-oscillator}
\end{figure}

The oscillator strength, defined as
\begin{equation}
F=\frac{|\int d^2 \bm{r}\Psi(\bm{r},
\bm{r})|^2 }{\int d^2\bm{r}_e \int d^2\bm{r}_h |\Psi(\bm{r}_e,\bm{r}_h)|^2}\ ,
\label{os}
\end{equation}
is plotted in Fig.~\ref{fig-oscillator}. Large values of $F$ corresponds to a
large transition matrix element from the exciton ground state into the vacuum.
We find from Fig.~\ref{fig-oscillator} that the results for large $\gamma$ are
 in good agreement with the 1D results.~\cite{RomR00} And when decreasing
the radius-to-width ratio $\gamma$, the value of $F$ does not suddenly drop to
zero, again emphasizing the  robustness of the excitonic AB effect in ring of
finite width.


\section{Conclusions}

Our results suggest that the excitonic AB effect originally predicted for a 1D model~\cite{Cha95,RomR00} remains essentially unchanged when allowing for rings of finite widths
as given by \eqref{eq-confiningpotential}. We find that when we enlarge the ring width by one order of magnitude from $1/\gamma=1/10$ to $1$, the magnitude of the AB oscillations drops
by about $15\%$ only. In addition, we show that the qualitative behavior of the oscillations both for the spectral position as well as the oscillator strengths of the exciton luminescence
lines are again governed by the relative strength of attractive Coulomb interaction to ring radius. Our results are in good agreement with recent experimental observations where the magnitude
of the excitonic AB oscillations was observed to be about $0.5\,$meV at binding energies of $4.35\,$mV for rings of about
$11$--$22\,$nm radius and $\gamma\approx 1$.~\cite{TeoCLM10}
We also note that our confining potential \eqref{eq-confiningpotential} has been chosen to retain its \emph{non-simply connectedness} due to the infinitely
repulsive centrifugal core at the centre. Hence even for very wide rings, there is an essential difference with respect to the previously considered 2D confining
potentials.~\cite{HuLZX01,HuZLX01,SonU01,GalBW02,GroGZ06,DaiZ07} This demonstrates that it is not so much the width or the exact shape of the  confining potential, but rather the avoidance
of the ring centre which is the important ingredient needed for the experimental observation of the excitonic AB effect.

Last, we expect that the effects of external electric fields~\cite{MasC03a,FisCPR09,LiP11} and the formation of charged excitons remain similarly robust in 2D,
whereas disorder effects~\cite{MeiTK01,HuZ03,DiaUG04} should be less important than in the 1D case.

\begin{acknowledgments}
We thank Andrea Fischer for valuable discussions and a critical reading of the manuscript. CGS is grateful to the Centre for Scientific Computing for hospitality and to
Ministerio de Educaci\'{o}n, Comunidad de Madrid and the European Social Fund for funding the research stays at Warwick during which much of this work was done. Work at Madrid was supported by
MICINN (projects Mosaico and MAT2010-17180).
\end{acknowledgments}

\bibliographystyle{prsty}\bibliography{bibliography/bibliograph}

\end{document}